# Colloidal Tin Sulfide Nanosheets:
# Formation Mechanism, Ligand-mediated Shape Tuning and Photo-detection

Fu Li,[1] Mohammad Mehdi Ramin Moayed,[1] Frauke Gerdes,[1] Sascha Kull,[1] Eugen Klein,[1] Rostyslav Lesyuk,[1,2] Christian Klinke[1,3,*]

[1] *Institute of Physical Chemistry, University of Hamburg, Grindelallee 117, 20146 Hamburg, Germany*

[2] *Pidstryhach Institute for applied problems of mechanics and mathematics of NAS of Ukraine, Naukowa str. 3b, 79060 Lviv, Ukraine*

[3] *Department of Chemistry, Swansea University - Singleton Park, Swansea SA2 8PP, United Kingdom*

* Corresponding author: christian.klinke@swansea.ac.uk

**Abstract**

Colloidal materials of tin(II) sulfide (SnS), as a layered semiconductor with a narrow band gap, are emerging as a potential alternative to the more toxic metal chalcogenides (PbS, PbSe, CdS, CdSe) for various applications such as electronic and optoelectronic devices. We describe a new and simple pathway to produce colloidal SnS nanosheets with large lateral sizes and controllable thickness, as well as single-crystallinity. The synthesis of the nanosheets is achieved by employing tin(II) acetate as tin precursor instead of harmful precursors such as bis[bis(trimethylsilyl)amino] tin(II) and halogen-involved precursors like tin chloride, which limits the large-scale production. We successfully tuned the morphology between squared nanosheets with lateral dimensions from 150 to about 500 nm and a thickness from 24 to 29 nm, and hexagonal nanosheets with lateral sizes from 230 to 1680 nm and heights ranging from 16 to 50 nm by varying the ligands oleic acid and trioctylphosphine. The formation mechanism of both shapes has been investigated in depth, which is also supported by DFT simulations. The optoelectronic measurements show their relatively high conductivity with a pronounced sensitivity to light, which is promising in terms of photo-switching, photo-sensing, and photovoltaic applications also due to their reduced toxicity.

**Introduction**

Two-dimensional (2D) metal chalcogenides serve as building blocks in various applications, including catalysis, batteries, solar cells, optoelectronics, and thermoelectric energy harvesting, owing to their unique size- and shape-dependent optical and electrical properties compared to 0D nanoparticles and also 1D nanorods.[1-8] 2D nanomaterials comprising lead or cadmium chalcogenides such as PbS, PbSe, CdS and CdSe have been intensively investigated. They exhibit tunable optoelectronic properties, which often are superior to their bulk counterparts.[9-12] However, they also display high toxicity, which hampers their applicability. To mitigate this problem, further studies of more environmentally friendly and less toxic semiconductors are necessary. One promising class of materials are tin chalcogenides.[13-15] In particular, tin(II) sulfide (SnS), a p-type semiconductor with an indirect bulk band gap of 1.07 eV and a direct band gap of 1.3 eV, shows both a high absorption coefficient and a good hole mobility.[16, 17] Typically, SnS adopts a layered orthorhombic (OR) crystal structure (*space group: Pbnm*). SnS can be described as highly distorted rock salt structure with atomic double layers, which are covalently bonded in the plane and van-der-Waals bonded in the vertical $b$ direction ($a$ = 4.33 Å, $b$ = 11.19 Å, $c$ = 3.98 Å). Latter is due to a chemically inert surface without any dangling bonds, which leads to its relatively high chemical stability and promotes 2D morphology.[18, 19] Recently, theoretical and experimental studies showed extraordinary properties of 2D SnS such as selected valleys optical excitation, ferroelectric and piezoelectric properties, which further motivate the development of cheap and safe methods for 2D morphology.[20, 21]

Recently, several studies focused on the synthesis of 2D SnS nanostructures. However, most of the materials are produced by demanding physical methods such as physical vapor deposition, molecular beam epitaxy and mechanical exfoliation, which usually yield polydisperse, irregular-shaped SnS nanosheets (NSs).[22-24] Compared to physical methods, colloidal syntheses can produce NSs from solution, assisted by surfactants to control the growth with uniform lateral dimensions, regular faceted edges and tunability. Such nanostructures can be processed by simple means, as there are spin coating and drop casting,[25] which is a prerequisite for efficient and inexpensive manufacture of super-assembled structures or deposition on flexible substrate for further applications.[26] However, most of the published synthetic recipes for SnS (analogically SnSe) yield mainly nanocubes and nanoparticles.[16, 27, 28] A few syntheses of colloidal 2D SnS nanosheets have been reported so far, but most of them are synthesized with halogen involved tin precursors (e.g. tin(IV) tetrachloride pentahydrate, tin(II) chloride) or with the assistance of hexamethyldisilazane (HMDS) for 2D nanomaterial formation.[16, 29-31]

The colloidal synthesis of 2D SnS NSs in the here presented work uses tin(II) acetate as tin precursor, which can be partly converted into the corresponding Sn-oleate complex during the conditioning step in the presence of oleic acid and partly serving as ligand for NS formation. In this process, no halogen ions are involved (e.g. through tin(II) chloride or tin(IV) chloride). Halogen ions in the SnS NS synthesis have been reported to be fully complexed by HMDS and be completely removed to avoid the inhibition of the formation of SnS NSs.[31, 32] Thus, the introducing of chloride ions is not necessary. The precursor bis[bis(trimethylsilyl)amino]tin(II) has been reported to yield rectangular NSs with a size of 7 μm × 20 nm.[33] However, this type of tin precursor is very flammable with low stability and high reactivity. Here, we introduce a halogen-and-HDMS-free synthesis of NSs, involving no flammable precursor as well. It shows that oleic acid (OA) and trioctylphosphine (TOP) amounts have strong control over shape and size (150-1680 nm). Based on experimental data and DFT calculations, we rationalize the shape changes of SnS NSs and propose an approach where the shape can be tuned between squared and hexagonal. The samples' crystal phase can be obtained as single-crystalline OR without other phases involved (e.g. pseudotetragonal phase from byproducts). We investigated in depth the mechanism of ligand-facet interaction to form NSs with different size and shape, as well as the function of the precursors. Eventually, by contacting individual NSs, their performance in terms of conductivity and photoconductivity has been investigated, revealing their outstanding optoelectronic properties.

**Experimental Section**

**Materials**. Tin(II) acetate (TA, 100%), oleic acid (OA, technical grade, 90%), trioctylphosphine (TOP, 97%), diphenyl ether (DPE, Reagent plus, ≥99%), toluene, dimethylformamide (DMF) were purchased from Sigma-Aldrich and were all used as-received without additional purification. Thioacetamide (TAA, ACS Reagent Grade, 99%) was bought from Fisher Scientific (Acros Organics). Thioacetamide, trioctylphosphine and tin(II) acetate were all stored in a glovebox.

**Synthesis of square-like 2D SnS nanosheets.** In a typical synthesis, a round-bottom three-neck 50 mL flask was used terminated by a condenser, a septum and a thermocouple. 59.2 mg (0.25 mmol) of TA, 0.2 mL (0.64 mmol) of OA and 0.5 mL of TOP (1.0 mmol) were dissolved in 10 mL of DPE. The mixture was degassed and dried under vacuum for 2 h at 75°C, partially transforming tin acetate into tin oleate and removing the free acetic acid from the system as well. After the vacuum step, the mixture was heated to 230 °C under nitrogen flow. After temperature stabilization (15 min), 19.5 mg (0.26 mmol) of TAA in 0.2 mL of DMF was injected rapidly into the reaction solution. After 5 min, the heating mantle was removed and the resultant solution was left for cooling to room temperature. The resultant nanostructures were then purified by centrifugation with toluene at 4000 rpm for 3 min (2-3 times). The product could then be re-suspended in toluene for further characterization or storage. To investigate the influence of different parameters, the other reaction parameters are kept constant.

**Characterization.** Transmission electron microscope (TEM) images and selected area electron diffraction (SAED) were obtained using a JEOL-1011 operated at 200 kV. All the TEM samples were prepared by dropping a 10 µL diluted toluene dispersion onto carbon-coated TEM grids followed by solvent evaporation at ambient conditions. The high resolution (HR) TEM images were performed on a Philips CM 300 UT microscope operated at 200 kV. The X-ray diffraction (XRD) patterns were obtained employing a Philips X'Pert System with Bragg-Brentano geometry, together with a copper anode at an X-ray wavelength of 0.154 nm. The samples were prepared by drop casting of a well suspended NS solution on silicon substrates. The surface topology data were obtained on an atomic force microscope (AFM) from JPK Instruments in intermittent contact mode. The samples were prepared by drop-casting a diluted NS suspension on a silicon chip. The samples were also measured by scanning electron microscopy (SEM) with a LEO GEMINI 1550 microscope for morphology information.

**Device preparation.** SnS NSs suspended in toluene were spin-coated on silicon wafers with 300 nm thermal silicon oxide as the gate dielectric. The highly doped silicon was used as back gate. The individual NSs were contacted by e-beam lithography followed by thermal evaporation of Ti/Au (1/55 nm) and lift-off.

**Device measurements.** Immediately after device fabrication, the samples were transferred to a probe station (Lakeshore-Desert) connected to a semiconductor parameter analyzer (Agilent B1500a). All the measurements have been performed in vacuum at room temperature. The vacuum chamber has a view port above the sample which is used for sample illumination. For illumination of the NSs, a red laser (627 nm, 1-16 mW) with a spot size of 2 mm was used.

**DFT simulations.** In order to evaluate the adsorption energies of the ligands on different crystal facets simulations based on density functional theory (DFT) were employed. For that the crystal geometry had been kept fixed to the experimental values for OR-SnS and the ligands were free to relax by geometry optimization. We used the versatile software package CP2K with the PADE LDA functional, the DZVP basis set, and a corresponding GTH-PADE potential. An individual SnS nanocrystal with 224 Sn and 224 S atoms and the respective ligand molecules are simulated with periodic boundary conditions where the box dimensions are sufficiently large to avoid interaction between virtual neighboring molecular structures.

## Results and Discussion

In the following we discuss the influence of various parameters on the synthesis of SnS nanosheets. It allows drawing conclusions on the formation mechanism and the crystallography. Eventually, we demonstrate their optoelectronic response.

**Synthesis of colloidal square-like SnS nanosheets.** TA, OA and TOP in DPE were degassed at 75°C and then heated to a reaction temperature of 230°C under nitrogen flow, followed by the hot injection of TAA, as sulfide precursor, to obtain fast nucleation and growth of 2D NSs. To guarantee the repeatability of the procedure, we kept the injection volume constant (0.2 mL) for all experiments. During the preheating and degassing phase, TA as tin precursor was partly transformed to tin oleate in the presence of OA. When tin (II) chloride was used as precursor, only irregularly shaped NSs and byproducts were formed with an otherwise unchanged recipe (Figure S1). A TEM image of square-like SnS NSs with a lateral size of approximately 460 nm is shown in Figure 1a. SAED (Figure 1b) reveals single-crystallinity due to the well-ordered dot pattern, which matches the list of reflections of OR-SnS bulk (The International Centre for Diffraction Data (ICDD) card 00-039-0354), respectively showing the diffraction planes (200), (101), and (002). The X-ray diffractogram (Figure 1c) of the capillary powder sample is also consistent with the crystallography for OR-SnS bulk. For the drop-casted SnS NSs film, the two pronounced peaks (040) and (080) are observed, representing the highly textured [010] orientation, which indicates that it is the one corresponding to the thickness. The Scherrer analysis of the (040) peak in the film sample yields a thickness of 29 nm for the sample shown in Figure 1c. A SEM image (Figure 1d) gives an overview on the morphology of the NSs, which shows the squared NSs with smooth surface and some squared NSs with truncated edges. An AFM image of the square-like SnS NSs is shown in Figure S2 and the measured thickness is 31 nm.

**Influence of trioctylphosphine on the synthesis of square-like single-crystal SnS nanosheets.** The synthesis of SnS NSs proceeds in two steps. First, the complexation of the tin precursor in the presence of OA happens before the injection of the second precursor. Second, the reaction between the tin and the sulfur precursor takes place to form nuclei, which is followed by the growth process to produce SnS NSs. Theoretically, the crystal facets with lower surface energy have a slower growth speed in order to minimize the total surface energy of the whole crystal based on the Gibbs-Curie-Wulff theorem and the Wulff construction.[34, 35] Furthermore, the reason for anisotropic growth of 2D nanocrystals is mainly due to the different surface energies of the facets according to the selective-adsorption of ligands. Thus, different growth speeds on each facet capped with certain ligands lead to the final shape of nanostructures. When there is no TOP and no OA in the synthesis, SnS nanoparticles and squared nanoplatelets (≤100 nm) are obtained (Figure S3). This indicates that the acetate can facilitate the formation of 2D nanostructures without TOP or OA in the synthesis, which is also supported by DFT simulations on the adsorption energy in Table 1. They show that acetate (AA$^-$) as ligand binds stronger to the (100) and (101) side facets, also binds strong to (010) facets, which could lead to truncated rectangular small nanoplatelets. However, it is shown that only acetate can mostly produce nanoparticles. When no TOP was involved in the synthesis, spherical nanoparticles and hexagonal NSs (Figure 2a, S3b) were synthesized with a constant amount of OA (0.64 mmol) and all other parameters unchanged. This is due to the strong bond of oleate on (100) and (101) (or {101}), which facilitate the hexagonal shape (elongated facet is (100), Figure S3b). A TEM image of the synthesis with 0.1 mmol TOP (Figure 2b) shows that most of the products were nanoparticles, accompanied by small hexagonal-shaped and square-shaped nanoplatelets. The appearance of square-shaped nanocrystals is consistent with the simulation data for TOP, revealing that adsorption energy on {101} facets are larger than others. Compared to the product with only OA ligands in Figure 2a, this indicates that this small amount of TOP still exerts its influence to maintain the 2D square-shaped nanostructure formation (~160 nm). More squared sheets with increased edge lengths of 240 nm were produced together with less irregular nanoparticles when 0.5 mmol of TOP was used (Figure 2c). 1.0 mmol TOP led to larger (nearly 460 nm) squared sheets without byproducts. Further, doubling the amount of TOP made no significant difference. Therefore, TOP plays an important role in retaining the growth of square-shaped NS with the acetate ligands, while OA facilitates the formation of hexagonal shaped nanostructures.

The XRD data (Figure 2f) display a more and more prominent (040) peak compared to other suppressed peaks with increasing TOP amount, which can indicate that more squared and larger 2D nanostructures as well as less byproducts (such as nanoparticles or nanoplatelets) were produced. The intensity of the diffraction peak at 44.04° is reduced when the amount of TOP increases from 0.0 mmol to 2.0 mmol TOP. This diffraction peak does not match the standard diffraction pattern of OR-SnS bulk but this peak is close to (220) of zinc blend (ZB) phase (43.8°). Theoretically, ZB SnS is a metastable phase, which is only kinetically stable and protected by a certain specific energy barrier avoiding the transformation to OR-SnS.[36, 37]

However, it can be better explained by the convolution of 43.9° and 44.1° of pseudotetragonal structure (PT), as more stable crystal phase under given condition.[16] The XRD measurement for the powder sample in capillary (Figure S4) demonstrates more clearly that the samples using 0.1 mmol of TOP show a mixture of OR and PT crystal structures, together with several small peaks (2Θ =32.7°, 35.5°) stemming from the planes (12$\bar{1}$), (031) of crystalline sulfur (ICDD card 01-072-2402). In addition, the peaks at 30.7°, 26.5° also stand for (101) and (021) of PT. The SAED also shows the nanosheet is OR-type and PT phase belongs to the byproduct (nanoparticle, Figure S5a, b). An increasing yield of crystalline SnS NSs could be obtained when we used TOP in amounts of up to 1.0 mmol in our case, with the gradually disappearance of PT particle product.

The thickness values calculated from XRD are in the range from 24 to 29 nm with a raising TOP amount (Figure S6). The SAED patterns of a single truncated square sheet from the synthesized sample with 1.0 mmol TOP (Figure 3) with two non-truncated diagonal corners and other two truncated corners are investigated. The HRTEM images (Figure 3b, c) and corresponding FFTs (inset) reveal the four planes corresponding to four sides of the square-shaped sheet, are {101} planes. This is also confirmed by the electron diffraction patterns in Figure 3d. The SAED pattern demonstrates the facets of (200), (002) and (101) which also are identified in the corresponding TEM images in Figure 3a (SAED patterns are rotated in respect to TEM image by the instrumentation). The truncated facets are found to be (100) and ($\bar{1}$00) facets. Figure 3e shows a model for SnS NSs with square shape (Figure 3e (1)), truncated square shape (Figure 3e (2)) and highly truncated or hexagonal shape (Figure 3e (3)). The atomic arrangements of the (101) and ($\bar{1}$01) facets (Figure 4a, b) show that Sn and S atoms are alternating ordered. In principle, each S (or Sn) atom in these facets has three bonds, from which one bond is always forming a bond with S (or Sn) perpendicular to the facets. As L-type ligands, TOP prefers to bind to metal centers. Our DFT simulations (Table 1) show the interaction between TOP and Sn atoms on the (101) facet and also demonstrate that TOP binds to the {101} facets most strongly and it displays a weak preference on the (010) facet. Therefore, acetate ligands lead to squared 2D SnS nanoplatelets and TOP can help to maintain the square shape and enlarge the size of the nanoplatelets to large nanosheets.

**Influence of the oleic acid amount in the synthesis of square-like single-crystal SnS nanosheets.** When only TOP and no OA is introduced in the synthesis, small squared sheets and nanoparticles both appear in the product (Figure 5a, Figure S8a). This matches the simulation data for TOP, indicating that TOP can enhance the formation of (101)-facet square NSs (Figure S8a). Further, we investigated the influence of the OA amount on the morphology of synthesized SnS NSs, keeping TOP (1.0 mmol) and other parameters constant. The TEM images of the synthesis using 0.32 mmol to 0.64 mmol of OA in Figure 5a-c show a lateral size change from 180 nm to 380 nm and thickness increase from 16 nm to 29 nm (derived from XRD, no shape change, still square-like sheets), together with a higher degree of uniformity in lateral-size. Moreover, squared NSs, as well as nanoparticles were observed with a low amount of OA (OA ≤ 0.64 mmol, TA: OA ratio ≤ 1: 2.5). This is due to that the strong adsorption energy of TOP on (101) dominates (Table 1), which assist the formation of squared sheets with the original acetate in the synthesis. The simulations are performed using simplified molecules for OA and oleate (represented by butyric acid BA and butyrate BA$^-$). The hexagonal shape starts to appear when the OA amount is larger than 2 mmol, which we consider as the minimum amount of OA for the formation of hexagonally shaped sheet. Lateral size and thickness values both increase largely (33 nm to 53 nm) when 12 times more OA than TA is applied (OA ≥ 3.2 mmol, TA: OA ratio ≥ 1: 12). Based on the law of mass action, the higher the amount of OA, the higher the possibility for OA to bind to the surface of the crystals, substituting the oleate. OA (L-type ligand) is a weaker ligand than oleate (X-type ligand), which facilitates the monomers to react on the crystal surface compared to the case of oleate as ligands, helping to grow larger and thicker sheets under otherwise unchanged conditions. Higher OA amounts (≥ 3.2 mmol) trigger the shape change from square-like to hexagonal nanostructures and strongly influence the size (TA: OA ratio ≥ 1: 12, especially 1:25). Analysis of the lateral sizes calculated from TEM images (Figure 5) shows a dramatic increase (from 160 nm to 1600 nm, OA used from 0.32 mmol to 6.34 mmol). This confirms that OA plays a major role in the size and shape tuning from squared to hexagonal SnS NSs. This is due to the largest adsorption energy of neutral OA ligands (represented by butyric acid BA, Table 1) on (100) facets compared to those on (101) or (010) facets, facilitating the formation of hexagonal NSs with elongated edges when there is an excess of OA ligands (≥ 3.2 mmol) (Figure S8b, with 6.4 mmol OA). XRD measurements show a further pronounced (040) peak and a more suppressed (220) peak for higher OA amount, indicating that a suitable quantity of OA could lead to stable OR-SnS nanostructures. Size and shape evolution of SnS NSs is sketched in Figure 6 and shows how OA with TOP influences the formation of SnS NSs.

**Effect of the precursor amount.** The amounts of precursors will define the numbers of nuclei for further growth. We found that with an increasing amount of tin precursor, there is a dramatic drop in size (Figure 7). When 0.025 mmol of TA is used, the lateral size reaches 826 nm, along with 36 nm of thickness (Figure 7). In the case of lower TA amount (<0.25 mmol), larger and thicker sheets are formed after the injection of sulfide source. The less the amount of tin source, the lower the number of nuclei that form, which then causes the formation of larger and thicker NSs. When the amount reaches 0.5 mmol TA, nanoparticles appear due to an excess of acetate ligands. Different amounts of sulfur source are also investigated to elucidate the influence on the SnS NS synthesis (Figure S9). Concerning the amount of sulfur precursor, the TEM images reveal that lower amounts (≤ 0.13 mmol) only lead to irregular shaped NSs. But these irregular sheets are still highly crystalline with OR crystal structure (Figure S10). This is attributed to the lack of sulfur monomers for further growth of NSs. XRD measurements also demonstrate that with an increasing amount of sulfur precursor, the (040) and (080) peaks are more pronounced and other peaks are more suppressed (Figure S9). When the amount of TAA reaches 0.26 mmol, lateral size and thickness begin to level off.

Therefore, we propose a growth mechanism for the formation of 2D SnS nanosheets based on our approach. Firstly, the tin ions are coordinated by acetate (Sn-acetate), as tridentate bridging ligands with certain chelate character, forming strongly distorted trigonal bipyramid for the first coordination sphere.[38] TAA thermally decomposes by hot injection, producing $S^{2-}$ for further reaction with $Sn^{2+}$ ions. Therefore, SnS nuclei are formed after TAA decomposition, passivated by the ligands on the surface of nuclei. With the injection of the sulfur source, the acetate ligands trigger the formation of SnS nanoparticles and squared nanoplatelets (≤100 nm) without OA and TOP in the reaction due to the dominated passivation on {101}, (100) and (010) facets by acetate. We introduce OA in the synthesis (0.64 mmol), hexagonal sheets and small nanoparticle byproducts are obtained, whereas square sheets are prepared with TOP involved (1.0 mmol) in the synthesis without OA, because of strong bond of oleate ligands on (100) and {101}. When OA is applied and the amount of OA is lower than 3.2 mmol together with 1.0 mmol TOP, Sn-acetate is partly replaced with Sn-oleate during the vacuum step, leading to the formation of still squared SnS NSs (~460 nm) without byproducts. However, hexagonal NSs are formed when the OA amount is higher with a constant TOP amount. Thus, the shape of the NSs is due to the balance between TOP and OA, as well as the original acetate ligands. For that, tuning the amount of TOP and OA can tune the shape between squared and hexagonal NSs. When the right ratio of TOP/OA is used in the synthesis, the product can be obtained as pure OR-SnS nanosheets without PT-type byproducts.

**Electrical measurements.** In order to investigate the potential application of the nanosheets, we measured the conductivity and photoconductivity of individually contacted SnS NSs. Figure 8a shows the I-V characteristics of the NSs in dark and under different illumination powers. The dark conductivity of the NSs is relatively higher than comparable materials such as colloidal PbS nanosheets (7.9 S/m, which is up to one order of magnitude higher than the conductivity of PbS nanosheets).[39] By illuminating the sheets with a red laser ($\lambda$ = 627 nm), the current increases, which is attributed to an increased carrier concertation due to the optical excitation and generation of electron-hole pairs. The sensitivity of the device, defined by *($I_{ill}$ − $I_{dark}$)/$I_{dark}$* reaches 0.95. Further, we calculated the spectral response $R_\lambda$ and detectivity $D^*$ by employing the following equations.

$$R_\lambda = \frac{I_{ph}}{P_{light} A}$$

$$D^* = \frac{R_\lambda A^{0.5}}{(2e I_{dark})^{0.5}}$$

In these equations, $I_{ph}$ is the photogenerated current, $P_{light}$ is the laser power density (50 mWcm$^{-1}$), $A$ is the effective area of the device (0.5 μm$^2$), and $e$ is the elementary charge.[46] The photodetectors based on our NSs show to have a spectral response of ~ 3×10$^3$ AW$^{-1}$ and a detectivity of ~ 4×10$^9$ Jones. The performance of these NSs is superior compared to the previously reported works on this material under similar conditions.[40] This is another proof for the high quality of the crystals and its low defect density.

By increasing the beam power, higher photocurrents are detected which is due to higher amounts of electron-hole pairs. However, as can be seen in Figure 8b, the dependence of the photoconductivity to the beam power experiences saturation in higher intensities. With an intermittent illumination, we also demonstrate the stability of the photocurrent. Similar values are achieved for the photoconductivity under illumination with 25 mW. Fast transitions between the on state and the off or dark state are observed (Figure 8c), which is an important requirement for photo-detectors. This is a further indicator for the high quality of the crystals and the lack of defects.[41] Also, through such measurements, clear change of the photocurrent can be observed by changing the beam power, while stability and speed of the system remain unchanged (Figure 8d). These measurements illustrate the potential of the produced SnS NSs for cost effective non-toxic optoelectronics (e.g. photo-detectors).

**Conclusion**

A facile and simple colloidal method has been explored to synthesize large (150 nm-1680 nm, thickness from 16 nm to 50 nm) single-crystalline SnS NSs in the presence of oleic acid and trioctylphosphine as co-ligands together with two precursors, tin acetate and thioacetamide. Therefore, no metal halides (e.g. tin chloride) or flammable organo-metallic precursors are introduced. The two ligands (OA (also oleate), TOP) are discovered to play a critical role in tuning the shape and size of the SnS NSs in addition to the original ligand, the acetate. The development of the final product involves the steps of instantaneous nucleation and anisotropic growth by the optimized balance of involved ligands. The NSs undergo a shape change between hexagonal sheets and square-like sheets by tuning the ligand quantity (acetate, oleate, OA, TOP). The crystal phase can be optimized from PT (coming from nanoparticle byproducts) and OR crystal structure (coming from nanosheets) into single-crystalline nanosheets of OR structure only. The conductivity and photoconductivity measurements demonstrate their high potential for optoelectronic applications such as photo-sensors and photo-switches.

**Acknowledgments**

The authors gratefully acknowledge financial support of the European Research Council via the ERC Starting Grant "2D-SYNETRA" (Seventh Framework Program FP7, Project: 304980) and a China Scholarship Council (CSC) PRC. C.K. thanks the German Research Foundation DFG for financial support in the frame of the Cluster of Excellence "Center of ultrafast imaging CUI" and the Heisenberg scholarship KL 1453/9-2.

**Supporting information**

Further experimental data are presented. In particular, detailed TEM images on the shape and crystallographic structure of the nanosheets, AFM images to determine the height of the materials, XRD measurements to determine the crystallographic phase. Beyond, SEM results are shown.

**Figures**

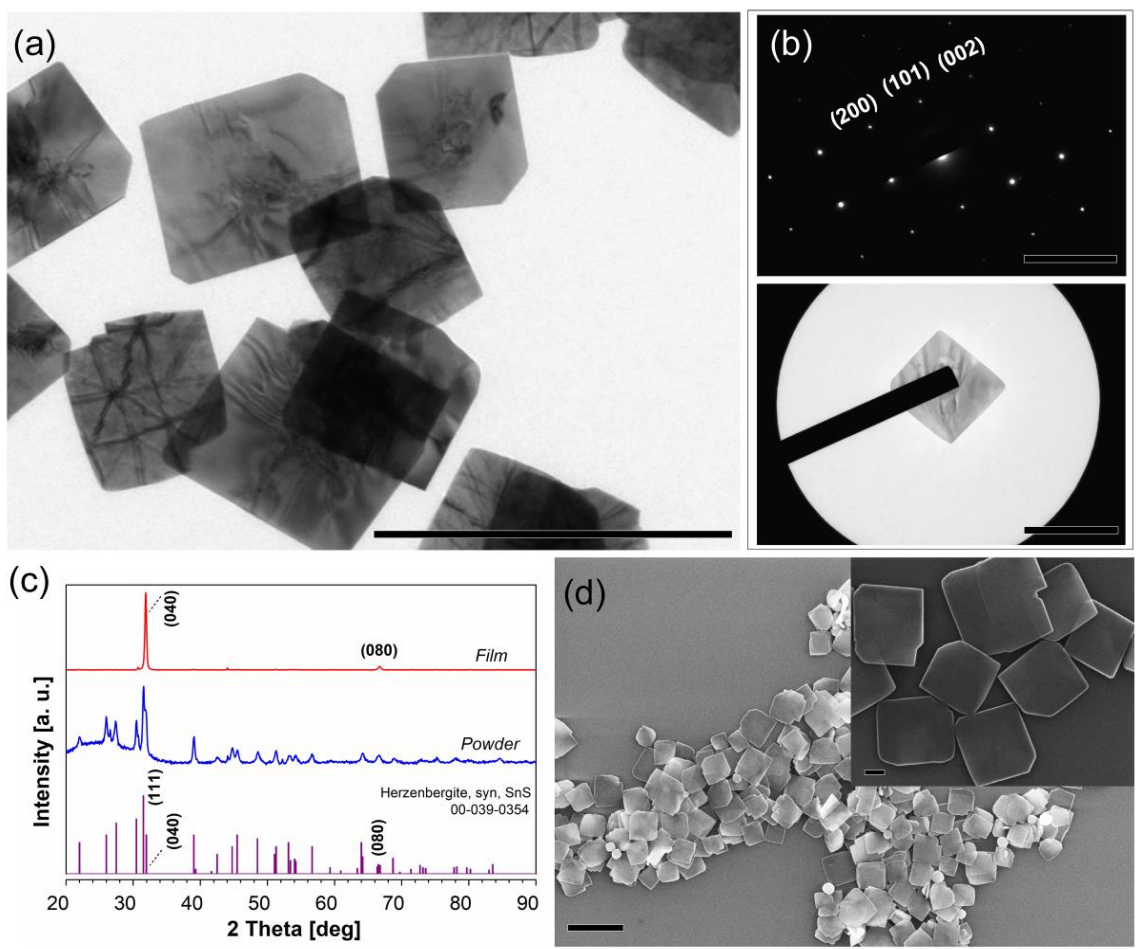

**Figure 1.** (a) TEM images of square-like SnS NSs synthesized with the standard synthesis. Scale bar =1 μm. (b) SAED pattern of a single nanosheet and corresponding TEM image. (c) XRD patterns for SnS NSs drop-casted on a Si wafer (thin film) and a powder sample in a capillary tube. (d) SEM images of SnS NSs (scale bar = 1 μm, inset: larger magnification: scale bar = 100 nm).

**Table 1.** Adsorption energy [eV] of ligand molecules on the (100), (101), (010) facets of SnS calculated by the density functional theory (DFT) method. The simulations were performed using the full version of acetate (AA⁻) and simplified molecules for TOP, OA and oleate (triethylphosphine (TEP) and tributylphosphine (TBP) for TOP, butyric acid (BA) for OA, and BA⁻ (butyrate) for the oleate). The simplified TOP, OA and oleate molecules were used to enable reasonable calculation times and to avoid additional contributions by the adsorption of the side chains. Anyhow, simulations on different chain lengths (C2 and C4) show that the tendencies are similar.

|  | SnS-101 side facet (isotropic) | SnS-100 side facet (anisotropic-zigzag) | SnS-010 top facet (Top or down) |
|---|---|---|---|
| TEP (C2) | 1.834 | 1.842 | 0.651 |
| TBP (C4) | 2.061 | 2.060 | 0.772 |
| AA⁻ (C2) | 3.661 | 4.453 | 2.506 |
| AA (C2) | 0.988 | 1.358 | 0.551 |
| BA⁻ (C4) | 3.635 | 4.355 | 2.547 |
| BA (C4) | 1.076 | 1.120 | 0.623 |

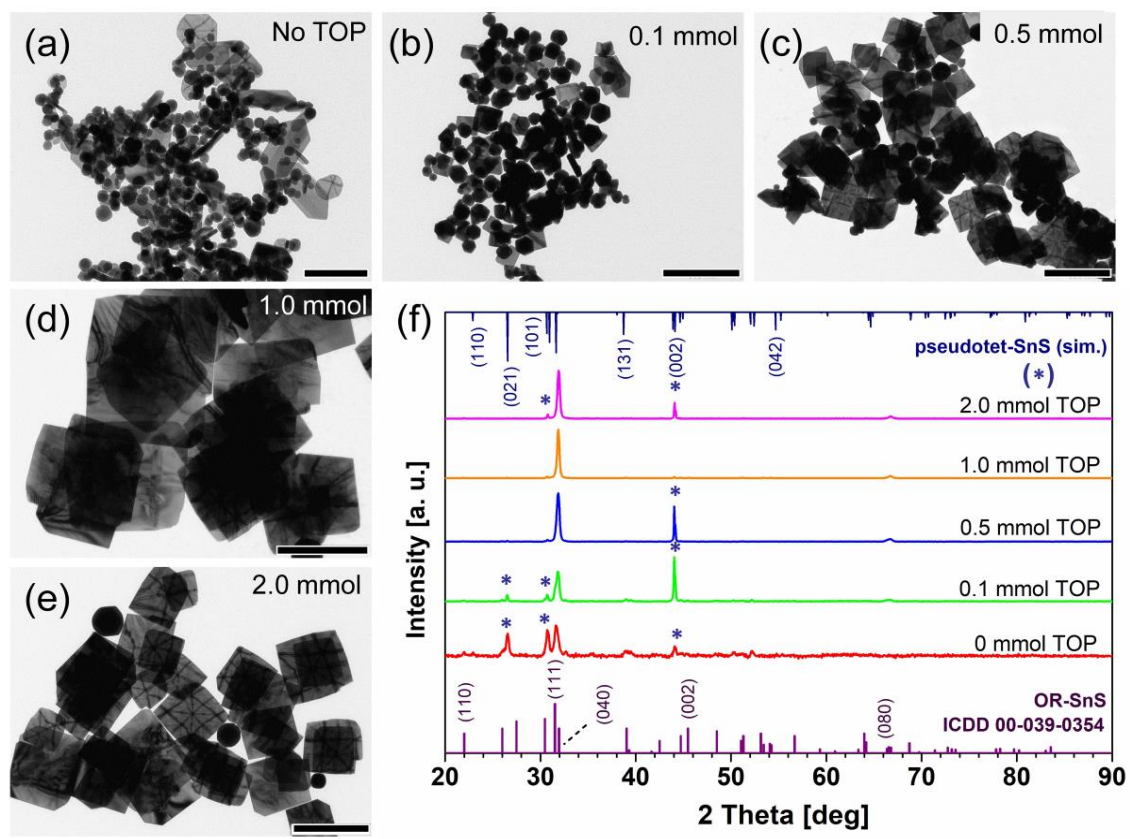

**Figure 2.** (a-e) TEM images of 2D SnS nanostructures with different TOP amounts (0 - 2.0 mmol). Scale bar is 200 nm for a, and 500 nm for b-e. (f) Powder XRD patterns of SnS NSs.

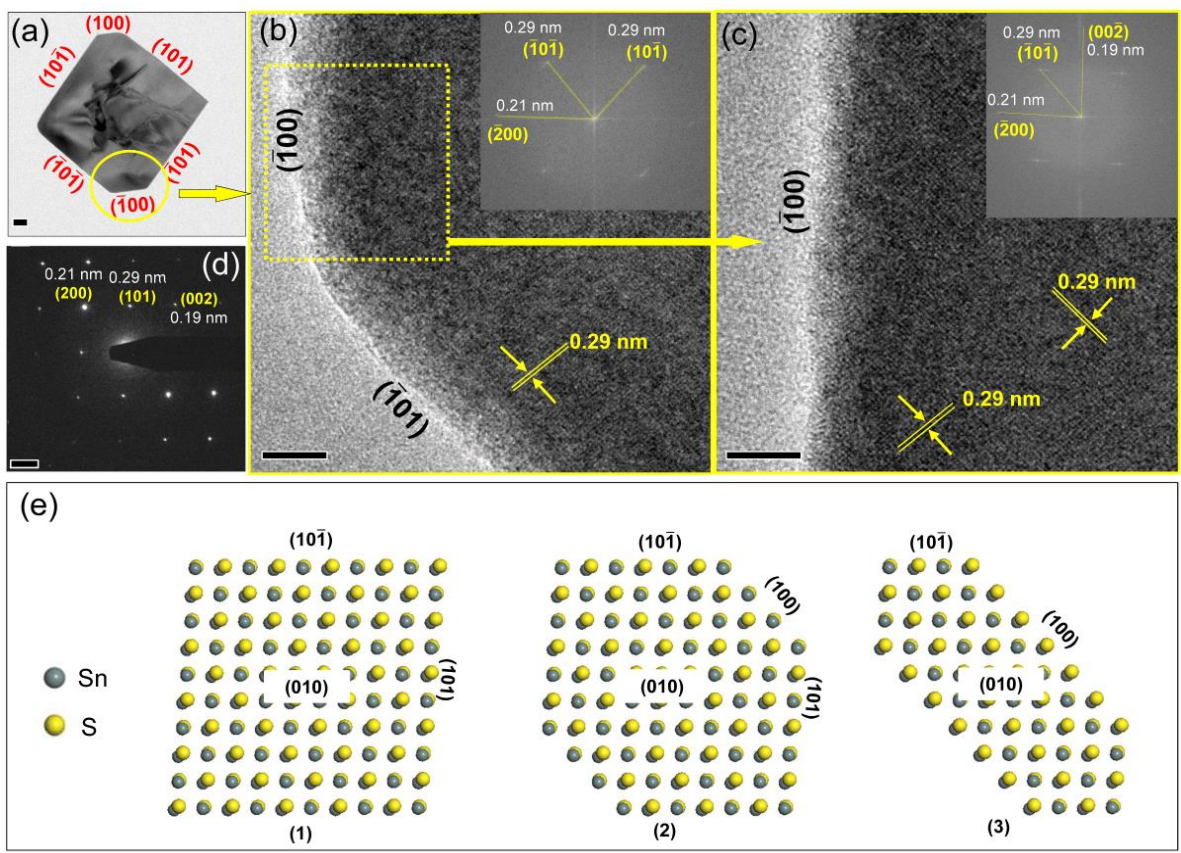

**Figure 3.** (a) TEM image of a single SnS NS from the standard synthesis with two non-truncated diagonal corners and other two truncated corners. The HRTEM image reveals the lattice spacing of 0.29 nm (b, c), which is consistent with the spacing of {101} planes. The SAED patterns (d) shows the lattice fringes of crystal facets (200), (002) and (101). The corresponding facets of ($\bar{1}$00), ($\bar{1}$01) are recognized based on the FFT analysis in inset of b and c, confirming the faceting in a. A set of atomic models of SnS NSs with the shape changing from squared to hexagonal (e). All these three models drawn here contained two layered atoms and that was why certain overlapping could be seen from the figure (grey Sn atoms overlap yellow S atoms, or yellow S atoms overlap grey Sn atoms). The scale bar = 50 nm in a, 5 nm in b and c, 5 μm in d.

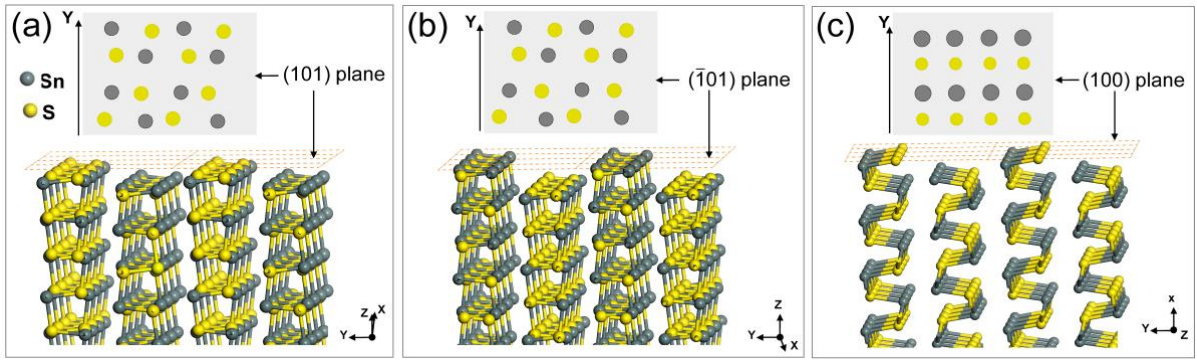

**Figure 4.** (a-c) The atomic arrangements in (101), ($\bar{1}$01) and (100) planes of SnS nanocrystals respectively. The top views of each model are also shown to demonstrate the exposed atoms on these three facets.

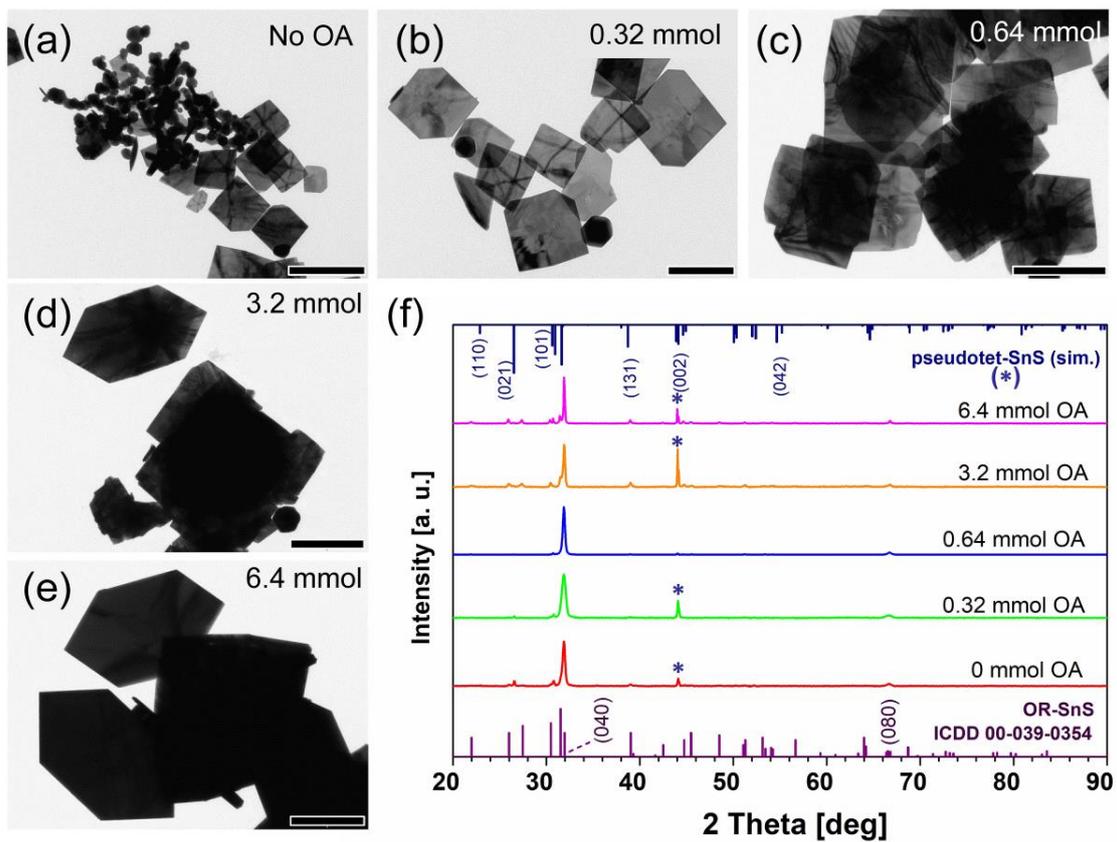

**Figure 5.** (a-e) Shape and size change of SnS nanoparticles to nanosheets synthesized using varied OA amounts (0 - 6.34 mmol). Scale bars correspond to 500 nm, 200 nm, 500 nm, 1 μm, and 1 μm respectively from a-e. (f) Powder XRD patterns of SnS NSs from a-e.

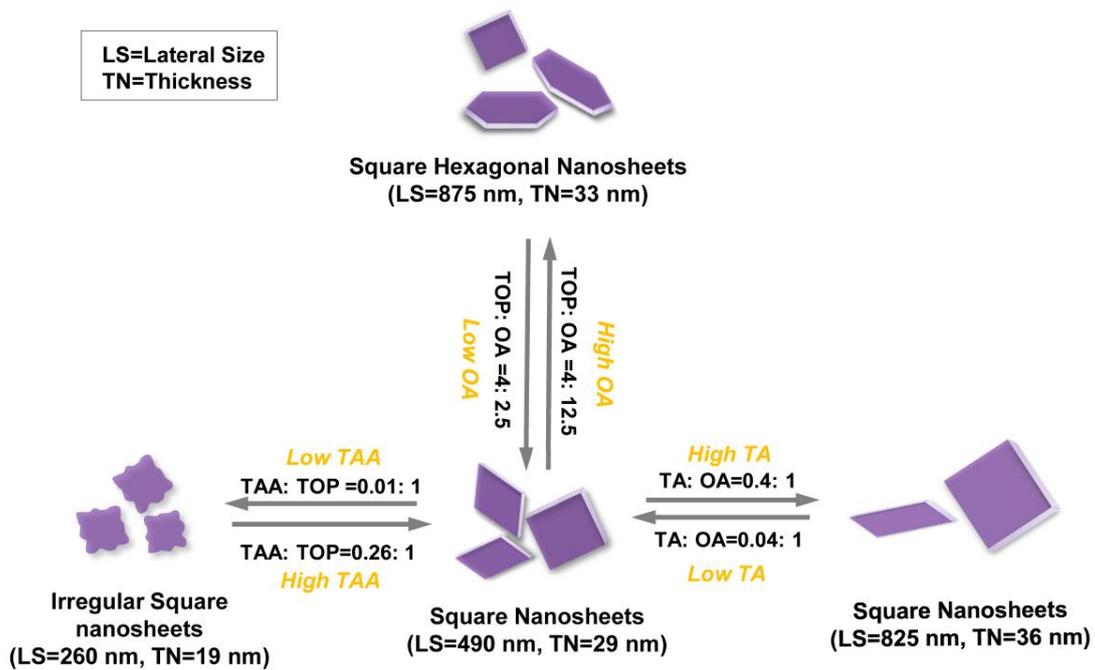

**Figure 6**. Schematic illustration of the formation and shape evolution of SnS nanostructures. LS=Lateral size, TN=Thickness.

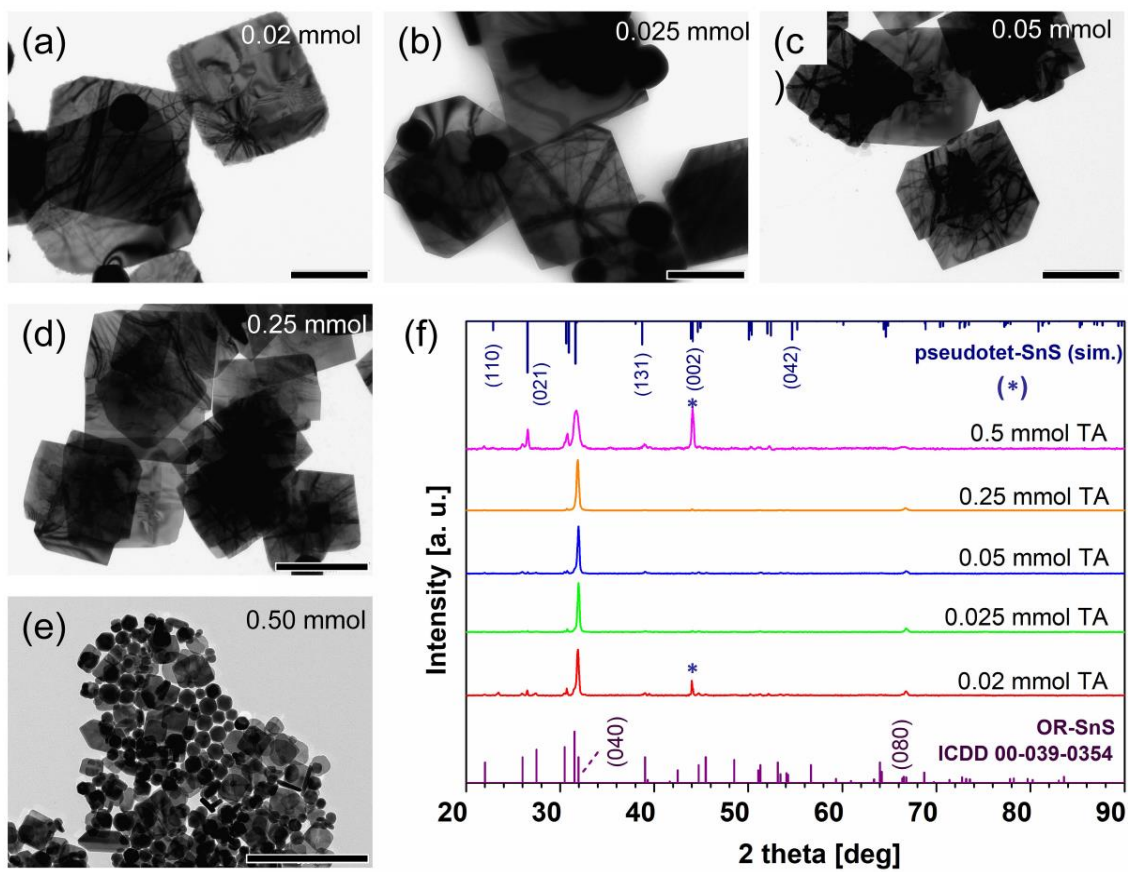

**Figure 7.** (a-e) Shape and size change of 2D SnS nanoparticles to nanosheets synthesized using an amount of tin acetate from 0.02 mmol to 0.025, 0.05, 0.25, and 0.5 mmol. Scale bars correspond all to 500 nm. (f) Powder XRD patterns of SnS NSs from a-e.

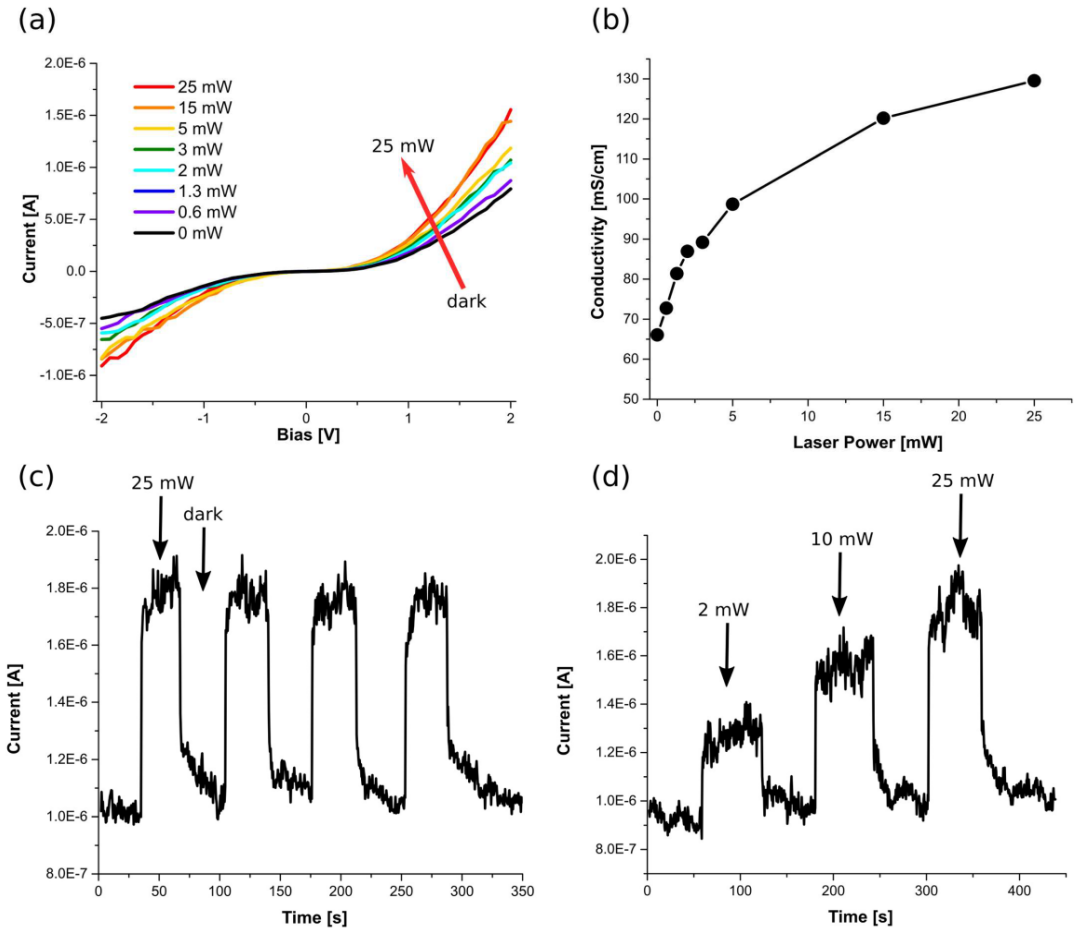

**Figure 8.** Photo-electrical properties of the nanosheets (a) I-V characteristics of the NSs in dark and under illumination with the red laser (λ = 627 nm) of various intensities. Higher currents are achieved under higher powers of illumination (b) Conductivity of the sheets in different laser powers. The photoconductivity increases by increasing the power and saturates in higher powers. (c) Stability of the photo-current under an intermittent illumination. By switching the 25 mW red laser on and off, stable photo-currents are achieved with fast transitions between the on state and the off state. (d) Intermittent illumination of the sheets with different powers. The photo-current can be tuned by the laser power, while the stability and the speed of the system are maintained.

**Graphic Abstract**

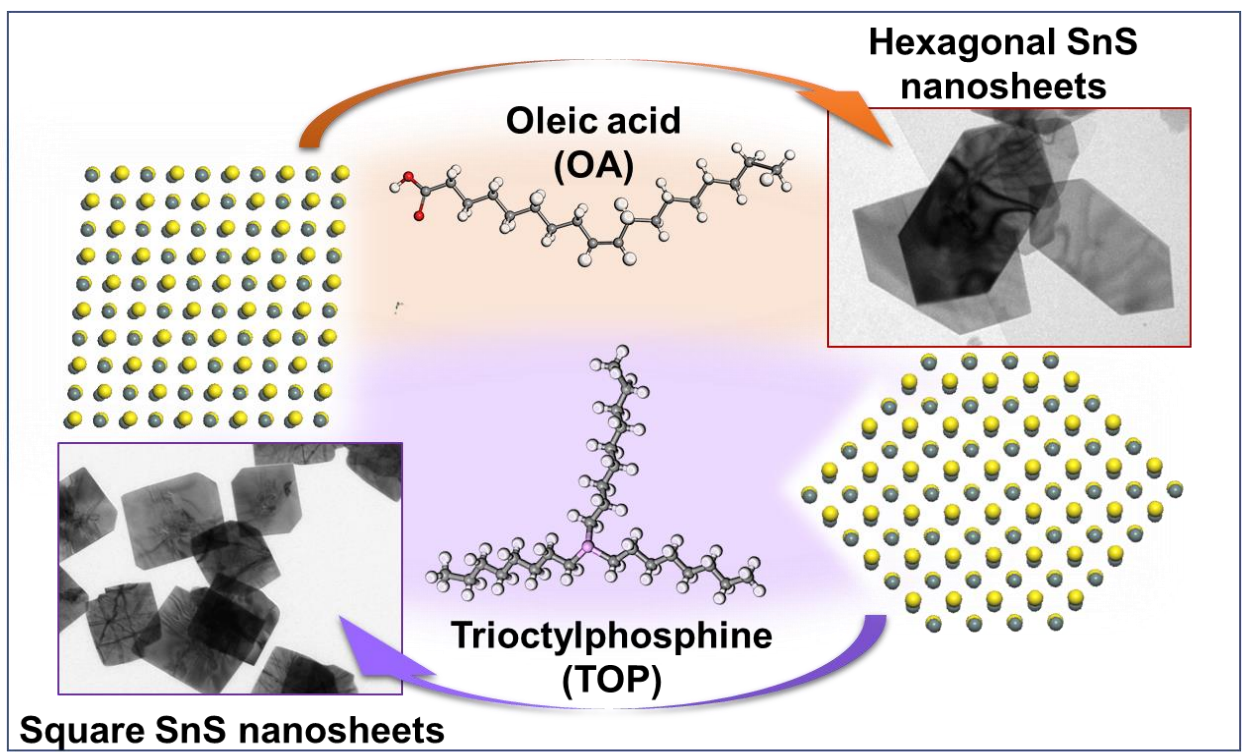

**Supporting information**

**Colloidal Tin Sulfide Nanosheets:**
**Formation Mechanism, Ligand-mediated Shape Tuning and Photo-detection**

Fu Li,[1] Mohammad Mehdi Ramin Moayed,[1] Frauke Gerdes,[1] Sascha Kull,[1] Eugen Klein,[1] Rostyslav Lesyuk,[1,2] Christian Klinke[1,3,*]

*[1] Institute of Physical Chemistry, University of Hamburg,*
*Grindelallee 117, 20146 Hamburg, Germany*

*[2] Pidstryhach Institute for applied problems of mechanics and mathematics of NAS of Ukraine, Naukowa str. 3b, 79060 Lviv, Ukraine*

*[3] Department of Chemistry, Swansea University - Singleton Park,*
*Swansea SA2 8PP, United Kingdom*

\* Corresponding author: christian.klinke@swansea.ac.uk

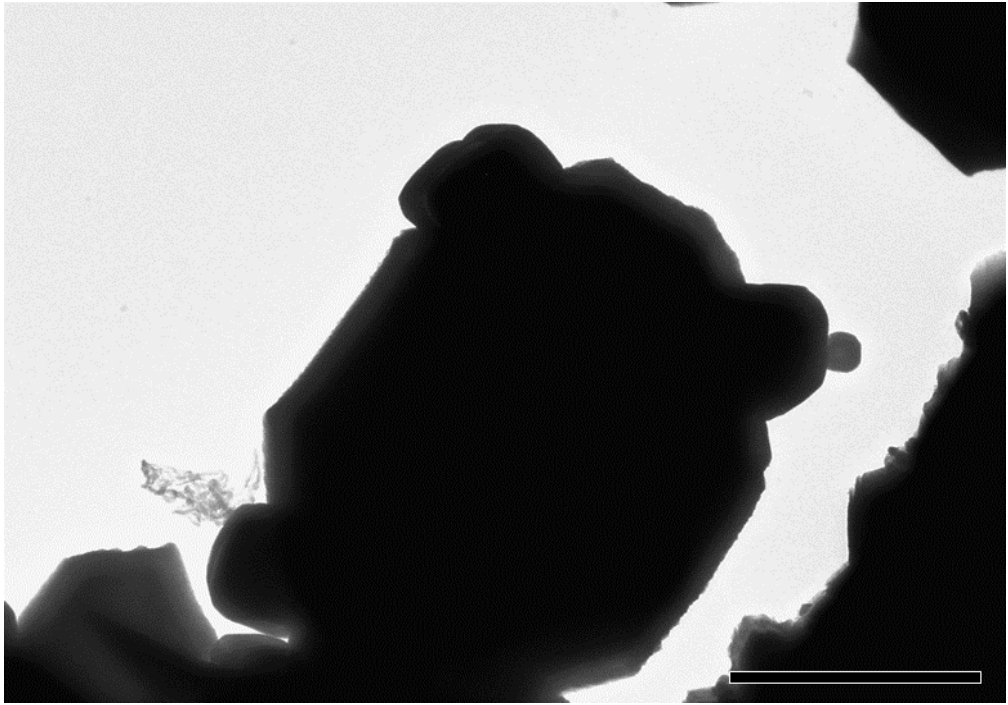

**Figure S1:** TEM image of SnS nanostructures synthesized with 0.25 mmol $SnCl_2 \cdot H_2O$ (scale bar = 1 μm).

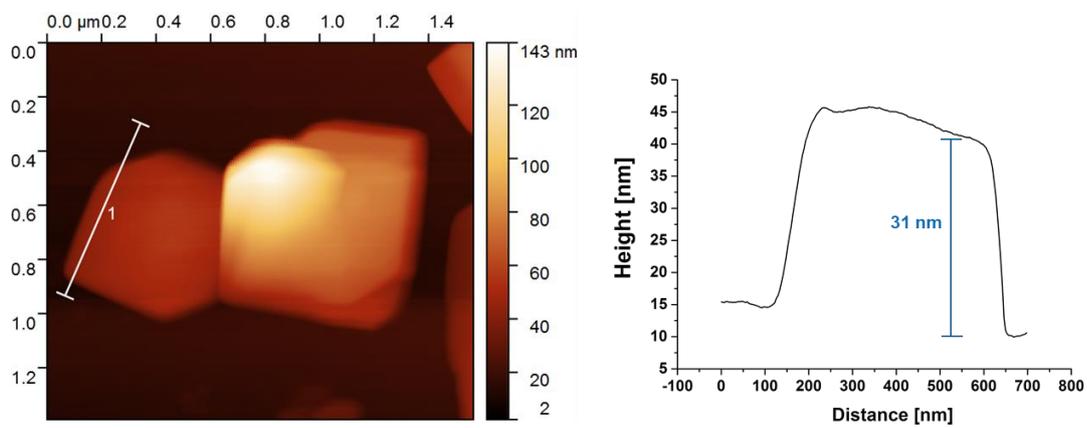

**Figure S2.** AFM image and the cross section of marked line 1 over a single SnS nanosheet demonstrated that the thickness was approximately 31 nm.

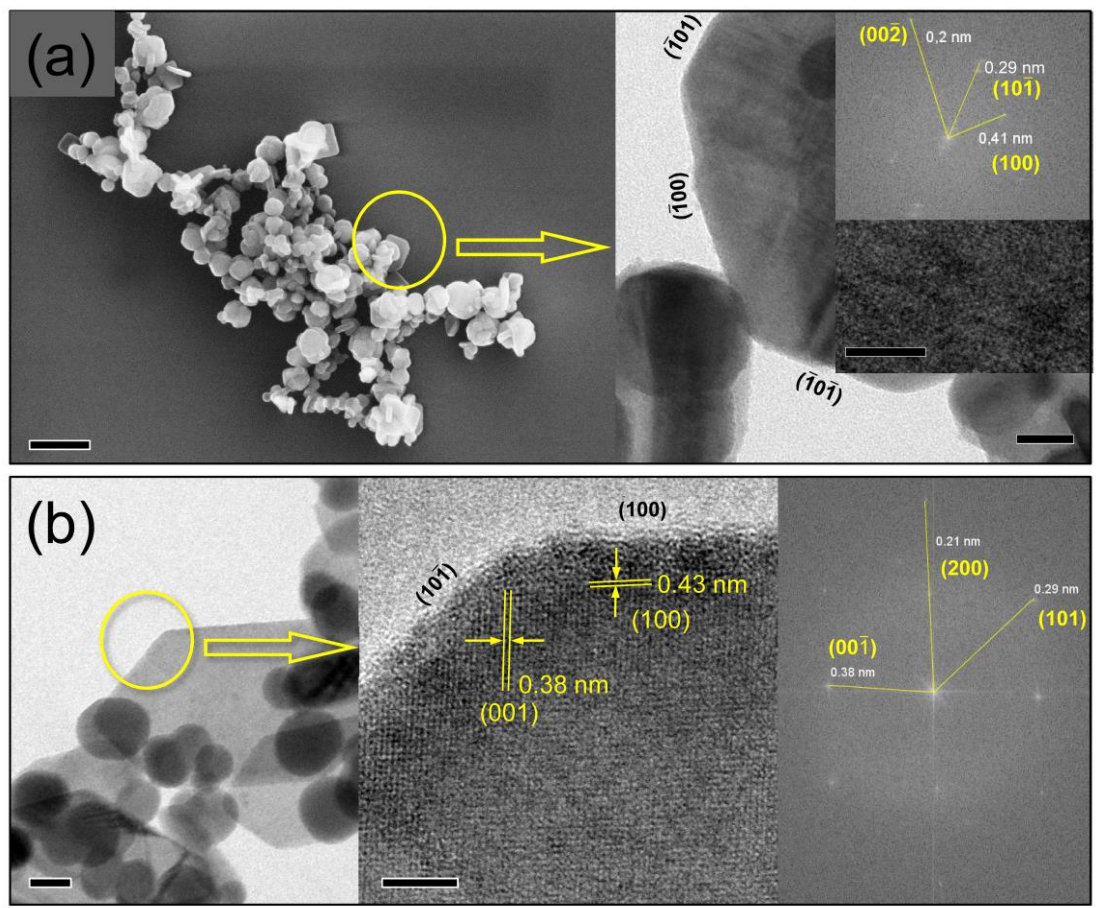

**Figure S3.** (a) SEM image, TEM image, HRTEM image, and resulting FFT of the synthesized nanocrystals without any TOP or OA ligands (Scale bar = 200 nm, 20 nm, 5 nm respectively). The small nanoplatelet shows a square shape with (100) and (001) side facets and (101) truncated facets. (b) TEM image, HRTEM image and resulting FFT of the synthesized hexagonal sheets with only OA (0.64 mmol) and no TOP (Scale bar=20 nm, 5 nm respectively). The sheet shows a hexagonal shape with elongated edges.

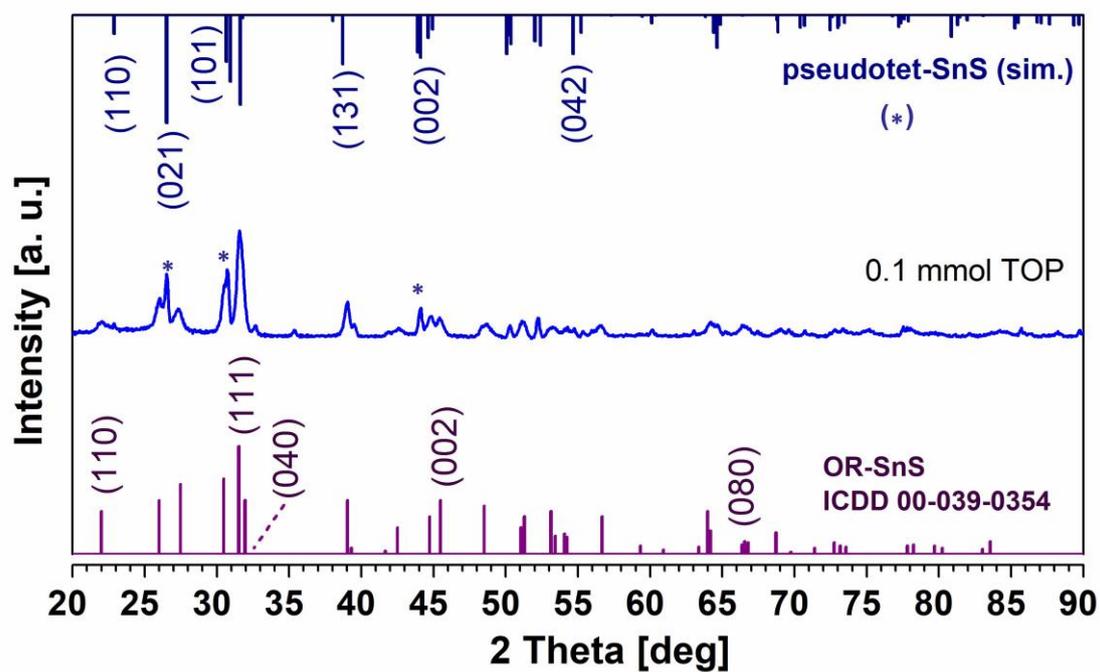

**Figure S4**. X-ray diffraction patterns (XRD) with the synthesized SnS nanostructures prepared in capillaries. This sample was produced by using 0.1 mmol TOP in the synthesis.

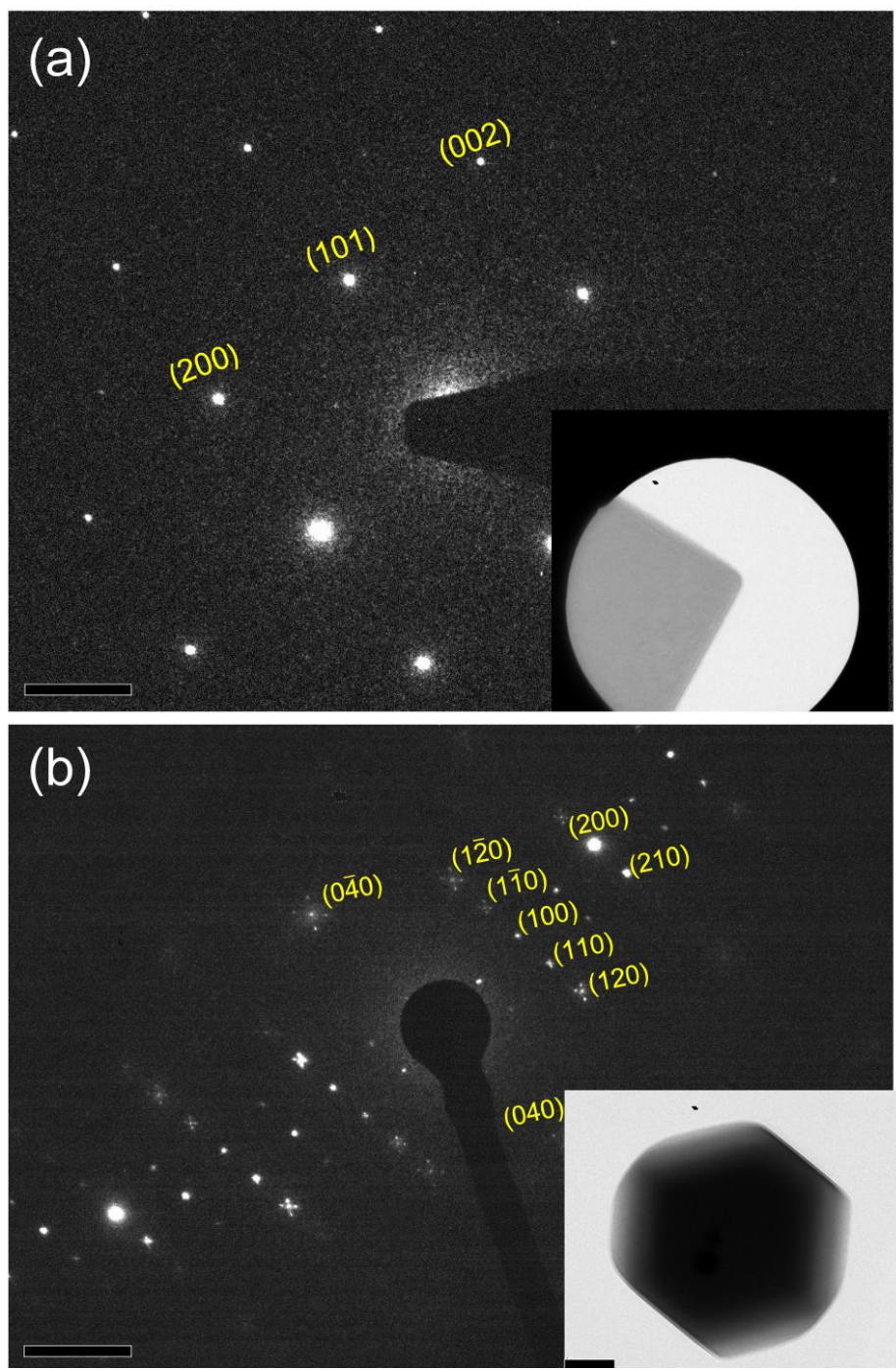

**Figure S5.** (a, b) Electron diffraction patterns of a single nanosheet (main product, Scale bar = 5 µm) and of a single nanoparticle (PT-type byproducts in the synthesis, Scale bar=2 1/nm, inset: Scale bar = 50 nm).

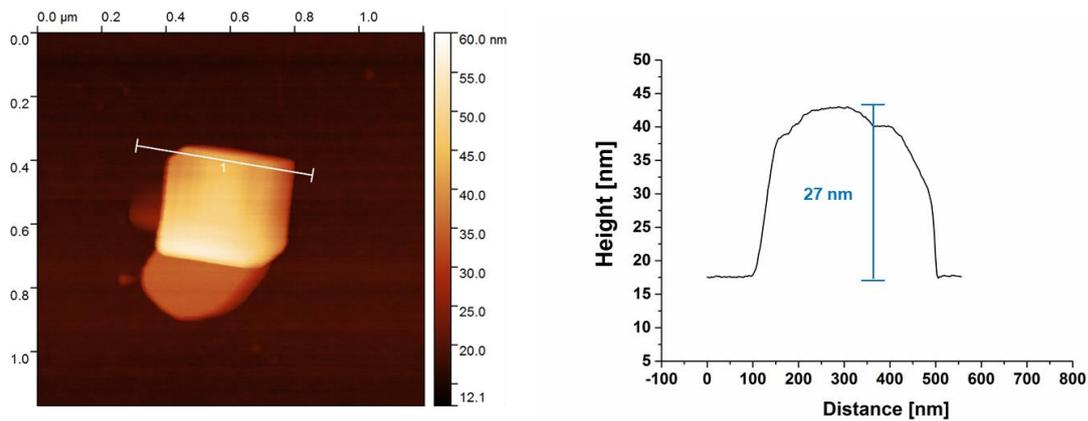

**Figure S6.** AFM image and the cross section of marked line 1 over a single SnS nanosheet demonstrated that the thickness was approximately 27 nm, which was slightly larger than the one measured based on XRD data (26 nm). The SnS NSs were synthesized with high TOP amount (2.0 mmol) with other parameters constant.

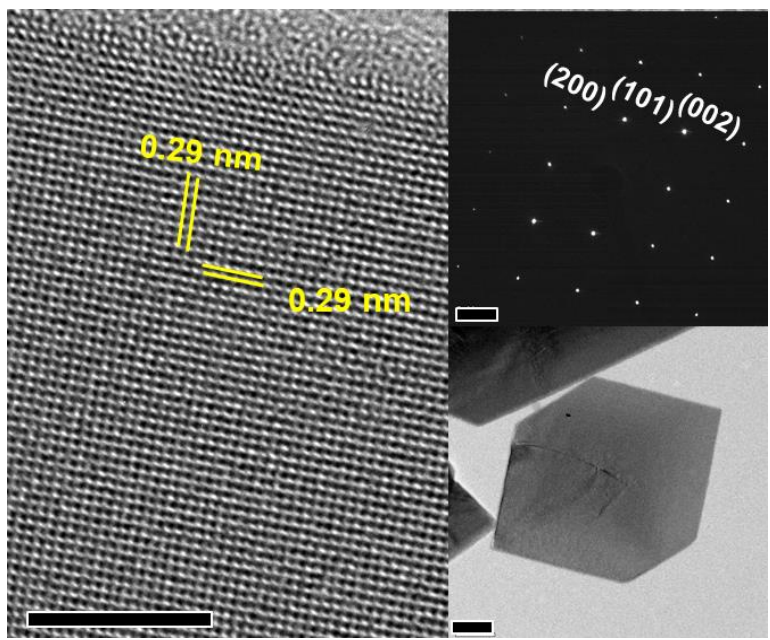

**Figure S7.** HR-TEM image of a single SnS nanosheet, together with measured lattice fringes (Scale bar=5 nm, inset: Scale bar=2 1/nm for the SAED pattern, and 50 nm for the TEM images).

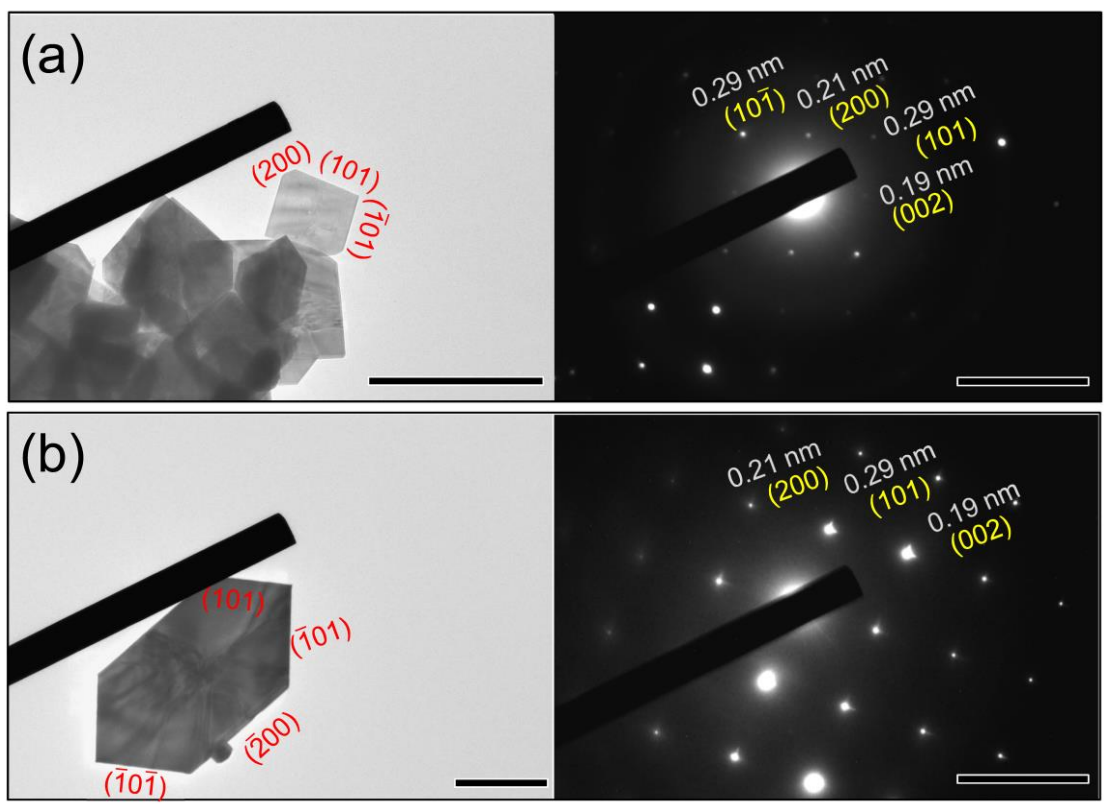

**Figure S8.** TEM images and SAED pattern of a square nanosheet synthesized with no OA and 1.0 mmol TOP (a), and a hexagonal nanosheet synthesized with 6.4 mmol OA and 1.0 mmol TOP (b). Scale bar=500 nm for both TEM images, Scale bar=20 μm for both SAED patterns. Rotations of SAED pattern in respect to TEM image have been performed.

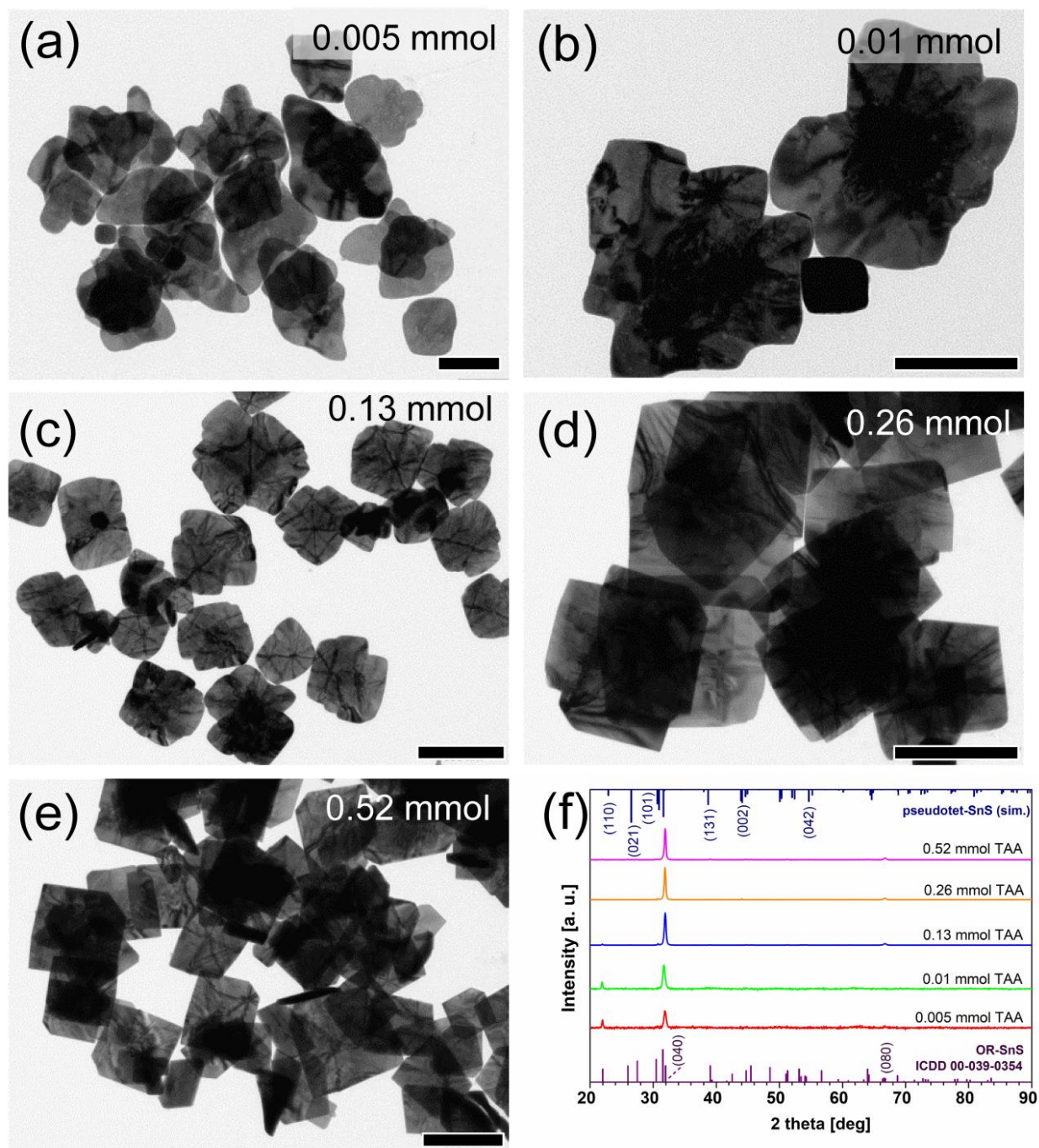

**Figure S9.** (a-e) Shape and size transformation of 2D SnS nanoparticles to nanosheets with TAA amount varied from 0.005 mmol to 0.01, 0.13, 0.26, 0.52 mmol. (f) Powder XRD patterns of SnS nanosheets from a-e. Scale bar = 200 nm, 200 nm, 500 nm, 500 nm, 500 nm respectively.

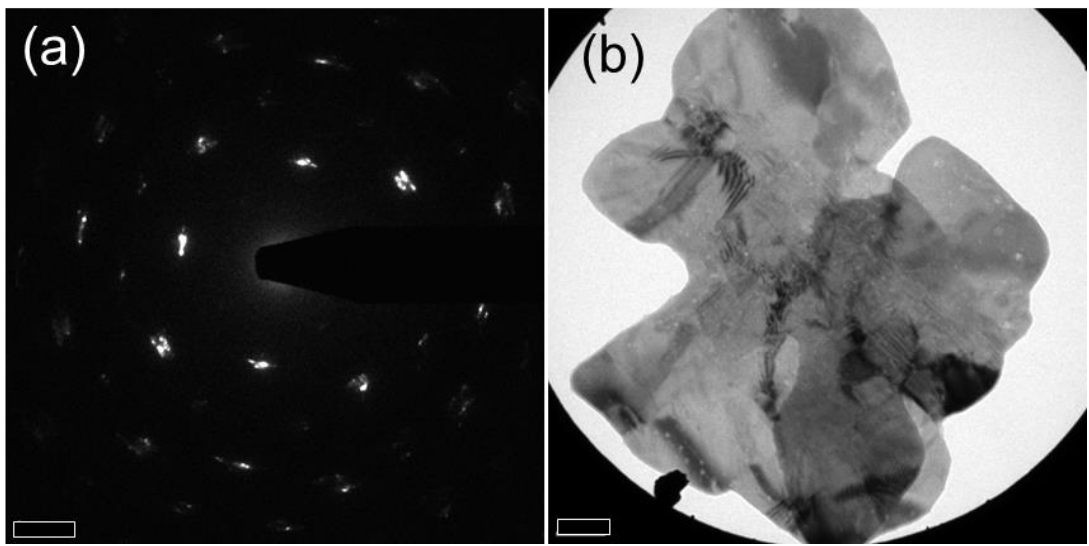

**Figure S10.** Electron diffraction pattern (a) and TEM image (b) and EDS analysis of the irregular shaped SnS nanosheets synthesized with 0.01 mmol TAA. Scale bar=5 μm, 50 nm for a, b.